\documentclass[12pt]{iopart}

\usepackage{iopams}
\usepackage[svgnames]{xcolor}
\usepackage[utf8]{inputenc}
\usepackage{dsfont}

\expandafter\let\csname equation*\endcsname\relax
\expandafter\let\csname endequation*\endcsname\relax

\usepackage{amsmath}
\usepackage{graphicx}
\usepackage{subfigure}
\usepackage{multirow}
\usepackage[colorlinks=true,linktocpage=true,hyperfigures=true,
            citecolor=blue,urlcolor=blue,linkcolor=blue]{hyperref}

\usepackage{units} 

\usepackage[utf8]{inputenc}
\usepackage[T1]{fontenc}
\usepackage{lmodern}
\usepackage{eurosym}

\usepackage{bm}
\usepackage{color}
\usepackage[colorinlistoftodos]{todonotes}

\def\Fig{figure~}

\def\Ref{ref.~}

\def\be{\begin{equation}}
\def\ee{\end{equation}}
\def\bea{\begin{eqnarray}}
\def\eea{\end{eqnarray}}

\def\ie{\textit{i.e.}~}
\def\eg{\textit{e.g.}~}

\def\keff{k_{\rm eff}}

\newcommand{\ket}[1]{\left| #1 \right\rangle}

\begin{document}

\title{Inertial quantum sensors using light and matter}

\author{B. Barrett$^{1,2}$, A. Bertoldi$^1$ and P. Bouyer$^1$}

\address{$^1$LP2N, Laboratoire de Photonique Num{\'e}rique et Nanosciences, Institut d'Optique Graduate School,
Rue Fran{\c c}ois Mitterrand, 33400, Talence, France\\
$^2$iXBlue, 34 rue de la Croix de Fer, 78100, Saint-Germain-en-Laye, France}
\ead{philippe.bouyer@institutoptique.fr}

\begin{abstract}
The past few decades have seen dramatic progress in our ability to manipulate and coherently control matter-waves. Although the duality between particles and waves has been well tested since de Broglie introduced the matter-wave analog of the optical wavelength in 1924, manipulating atoms with a level of coherence that enables one to use these properties for precision measurements has only become possible with our ability to produce atomic samples exhibiting temperatures of only a few millionths of a degree above absolute zero. Since the initial experiments a few decades ago, the field of atom optics has developed in many ways, with both fundamental and applied significance. The exquisite control of matter waves offers the prospect of a new generation of force sensors exhibiting unprecedented sensitivity and accuracy, for applications from navigation and geophysics to tests of general relativity. Thanks to the latest developments in this field, the first commercial products using this quantum technology are now available. In the future, our ability to create large coherent ensembles of atoms will allow us an even more precise control of the matter-wave and the ability to create highly entangled states for non-classical atom interferometry.
\end{abstract}


\section{Introduction}

For nearly one century, quantum physics did not cease to reveal surprising, intriguing and sometimes wonderful phenomena. It allows, for example, to reconcile the wave and corpuscular descriptions of light that are contradictory in classical physics, and formed one of the great controversies of the end of the 19$^{\rm th}$ century. Since 1924, when Louis de Broglie generalized to massive particles the wave-particle duality highlighted by Plank and Einstein for photons \cite{deBroglie1923}, we have been able to define a wavelength for matter---the so-called de Broglie wavelength $\lambda_{\rm dB} = h/p$, which is related to a particle's momentum $\bm{p} = m \bm{v}$. Shortly afterwards, the first matter-wave diffraction experiments were carried out with electrons \cite{Davisson1927}, and later with a beam of He atoms \cite{Estermann1930}. Although these novel experiments opened the way toward \emph{atom optics} and interferometry with matter waves, they also revealed two major challenges. First, due to the relatively high temperature of most accessible particles, typical de Broglie wavelengths were much less than a nanometer (thousands of times smaller than that of visible light)---making the wave-like behavior of particles difficult to observe. For a long time, only low-mass particles such as neutrons or electrons could be coaxed to behave like waves since their small mass resulted in a relatively large de Broglie wavelength. Second, there is no natural mirror or beam-splitter for matter waves because solid matter usually scatters or absorbs atoms. Initially, diffraction from the surface of solids, and later from micro-fabricated gratings, was used as the first type of atom optic. After the development of the laser in the 1960's, it became possible to use the electric dipole force to interact with atoms using coherent radiation.

\begin{figure}[ht]
  \centering
  \includegraphics[width=0.5\textwidth]{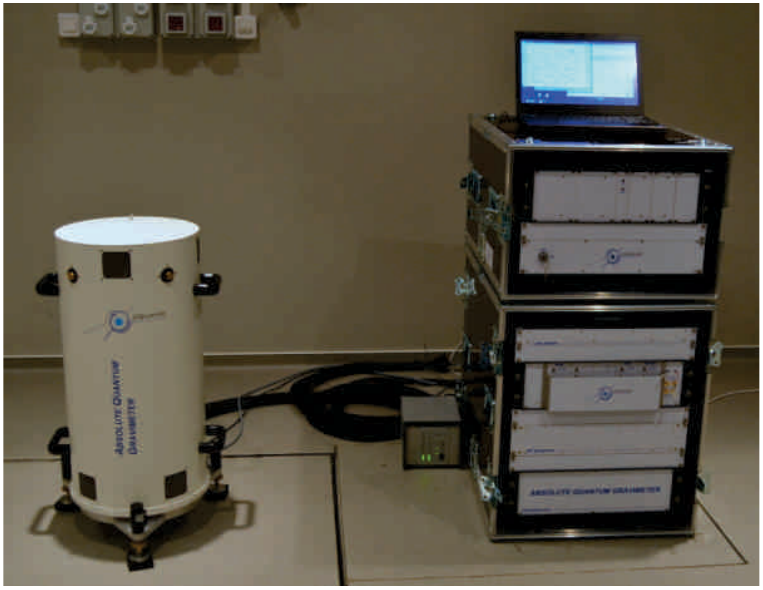}
  \caption{Photography of the Absolute Quantum Gravimeter available at Muquans. Courtesy of B. Desruelle -- \texttt{http://www.muquans.com}}
  \label{fig:Muquans}
\end{figure}

During the early 1990's, the physicists began to manipulate atomic de Broglie waves and the field of atom optics emerged \cite{Berman1997, Cronin2009}. The first experiments diffracted atomic matter-waves using micro-fabricated material gratings \cite{Keith1988}, ``light'' gratings from interfering laser fields \cite{Moskowitz1983, Kasevich1991a}, and evanescent waves from a high-reflection coating on a glass surface \cite{Kasevich1990}. Using these atom-optical elements, it is possible to observe atomic de Broglie wave interference in \emph{atom interferometers} \cite{Borde1989, Keith1991, Kasevich1991b}, which have proven themselves as invaluable tools for the study of fundamental physics, high-precision measurements, and inertial sensing. For instance, it is possible to measure the acceleration of gravity with an accuracy of 1 part per billion (ppb) \cite{Farah2014}, the rotation of the Earth with an accuracy better than 1 millidegree per hour \cite{Gustavson1997, Barrett2014b}, or to detect minute changes in gravity caused by mass displacements \cite{Rosi2015} or ocean tides \cite{Peters2001}.  These devices are so precise that they are used today as references for fundamental constants (mass, gravity), and are powerful candidates to test the theory of General Relativity on surface-based \cite{Dimopoulos2007, Hartwig2015, Zhou2015}, subterranean \cite{Canuel2014} or in Space-based laboratories \cite{Altschul2015, Williams2016}. Projects are currently underway to verify the universality of free fall (UFF) \cite{Hartwig2015, Barrett2015, Zhou2015, Williams2016, Fray2004, Schlippert2014, Aguilera2014}, to detect gravitational waves in a frequency range yet unreachable with current laser-based detectors \cite{Dimopoulos2009, Geiger2015, Chaibi2016}, and to test dark energy \cite{Burrage2015, Hamilton2015}. Nowadays, many efforts are devoted to designing compact, robust and mobile sensors \cite{Barrett2014a, Freier2015}. The ability to capture and immobilize the particles in traps \cite{Ashkin1978, Phillips1985, Chu1997}, where the atoms exhibit velocities of a few millimeters per second, has lead to a new generation of atomic sensors that are operated in aircraft \cite{Geiger2011} and soon in rockets \cite{MAIUS2016}, that are commercially available (see \Fig \ref{fig:Muquans}) and could be the next generation of navigation unit \cite{Canuel2006}.

Finally, the possibilities to obtain such low temperatures or even lower (hundreds of nanoKelvin, where the intrinsic quantum behavior of the atoms is even directly visible \cite{Anderson1995, Davis1995, Cornell2001}) offers new fascinating possibilities of reaching even more compact systems, or of using advanced quantum manipulation to increase the sensitivity of these sensors. Utilizing non-destructive measurements, for instance, can allow one to extend the interrogation time beyond the present limits and could lead to the preparation of highly entangled states leading to non-classical atom interferometry.

\section{Quantum sensor using light and matter}

\subsection{How a light pulse ``splits'' an atom}

The key element for an atom interferometer consists of performing a coherent splitting of the matter wave. If we consider a two-level atom (\ie an atom with a ground state $\ket{g}$ and an excited state $\ket{e}$) with initial position $\bm{r}$ and momentum $\bm{p}$, characterized by a wavefunction $\psi_i \sim \exp(i \bm{p} \cdot \bm{r}/\hbar)$, an atomic beam-splitter can be achieved by engineering an interaction potential which will evolve the wavepacket into a superposition of momentum states. One such interaction (the electric dipole force) can spatially modulate the amplitude of the wavefunction by $\sin(\bm{k} \cdot \bm{r})$ so that $\psi$ becomes $\sin(\bm{k} \cdot \bm{r}) \psi$. Decomposing $\sin(\bm{k} \cdot \bm{r})$ into a sum of exponentials immediately shows the final wavefunction $\psi_f \sim (e^{i \bm{k} \cdot \bm{r}} - e^{-i \bm{k} \cdot \bm{r}}) \psi_i$. This is a coherent superposition of two momenta, $\bm{p} + \hbar \bm{k}$ and $\bm{p} - \hbar \bm{k}$. The light-matter interaction can also be engineered to create a phase grating which spatially modulates the \emph{phase} of the wavefunction. In this case, $\psi_i$ becomes $e^{i \bm{k} \cdot \bm{r}} \psi_i$---resulting in a momentum translation $\bm{p} \to \bm{p} + \hbar \bm{k}$.

\begin{figure}[!tb]
  \centering
  \includegraphics[width=0.5\textwidth]{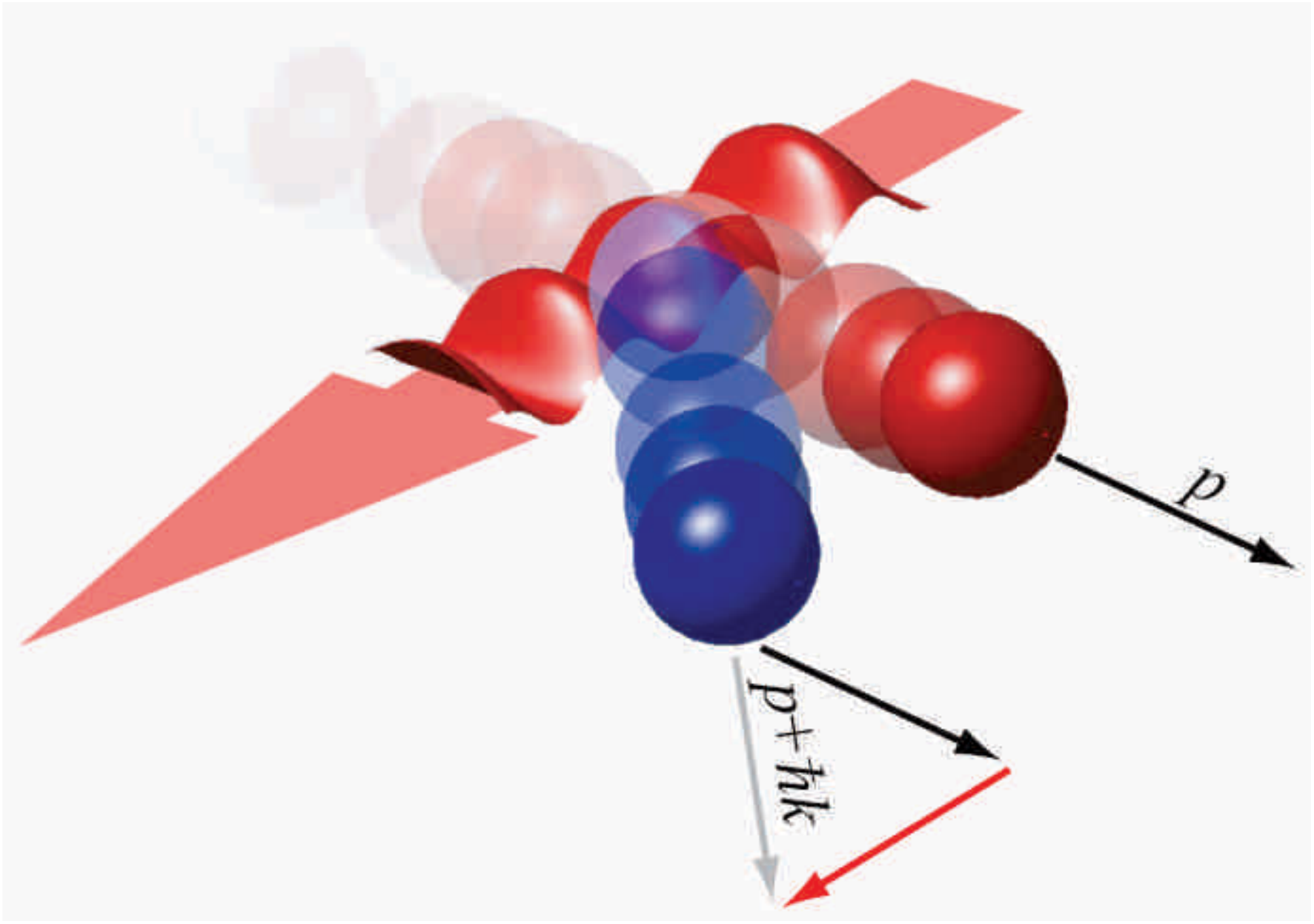}
  \caption{Atom wavepacket split by a pulse of light: the momentum carried by the light (red arrow) can be coherently transferred to the atoms, with a probability between 0 and 1. In quantum mechanics, this creates a coherent superposition of atoms with momentum $\bm{p}$, which did not absorb a photon (in red), and atoms with momentum $\bm{p} + \hbar \bm{k}$, which did absorb a photon (in blue).}
  \label{fig:BeamSplitter}
\end{figure}

These types of interactions are strongest when the light is nearly resonant with the transition frequency between the two internal states $\ket{e}$ and $\ket{g}$. When the optical phase grating is nearly resonant with the atom, the combined atom-field energy of the atomic states becomes spatially modulated $\delta E(r)$. For an interaction time $t$ of the atom with the field, the energy offset $\delta E(r)$ results in a phase shift $\delta \phi(r) t$. The interaction time is set, for example, by the time of flight of the atom through the field, or by the pulse duration of the laser field intensity. As for a Bragg diffraction grating in optics, the amplitude of the phase shift (\ie the time $t$ spent by the atoms in the diffracting field) will influence the fraction of the wavefunction $\psi_i$ actually translated by $\hbar \bm{k}$ and result in $\psi_f \sim \alpha\,\psi_i + \beta\,e^{i \bm{k} \cdot \bm{r}} \psi_i$. The interaction time can be chosen, for example, so that $\alpha = \beta = 1/\sqrt{2}$ to implement a beam splitter or so that $\alpha = 1 - \beta = 0$ to implement a mirror. In addition, the internal state of the atom becomes correlated with its external momentum. That is, atoms in the excited state must have absorbed a photon, ergo they must have a momentum $\bm{p} + \hbar \bm{k}$. This has significant practical advantages: the momentum state can be determined simply by detecting the internal state of the atom. In practice, two-photon stimulated Raman transitions \cite{Kasevich1991a} between ground state hyperfine levels have proven to be particularly fruitful for implementing this class of beam splitter.

\subsection{Light pulse matter-wave interferometer}

\begin{figure}[!tb]
  \centering
  \includegraphics[width=0.8\textwidth]{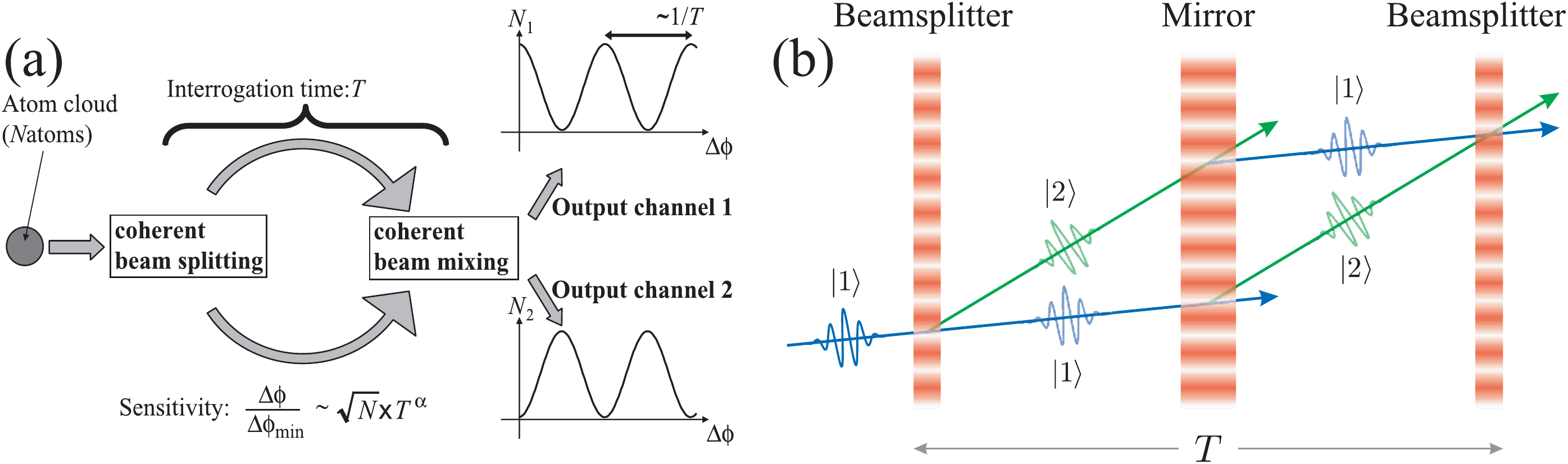}
  \caption{(a) A matter-wave is split into two parts by the first beam splitter. The wavepackets then propagate along the two different paths for an \emph{interrogation time} $T$, during which the they can accumulate different phases. At the last beam-splitter, the number of atoms at each output is modulated with respect to the phase difference accumulated over the two paths. (b) The Mach-Zehnder configuration uses a series of three pulses: the first pulse separates the atomic wave, the second redirects the two partial waves and the final pulse causes the two wavepackets to recombine and interfere. The interference is detected, for example, by measuring the number of atoms in one of the output states.}
  \label{fig:InterferencePrinciple}
\end{figure}

In general, an atom interferomer uses a succession of \emph{coherent} phase-locked beam-splitting processes separated by a time $T$ to an ensemble of particles, followed by detection of these particles in each of the two interferometer output channels. The interpretation in terms of matter waves follows from a direct analogy with photonic interferometry. The incoming wave is split into two different paths by the first beam-splitter. The accumulation of phase along the two paths leads to interference at the last beam-splitter, producing complementary probability amplitudes in the two output channels. The detection probability in each channel is then a sinusoidal function of the accumulated phase difference, $\Delta \phi$. Atom interferometers are generally based on the Mach-Zehnder design: two splitting processes with a mirror inserted inside to fold the paths (see \Fig \ref{fig:InterferencePrinciple}). Usually, the diffraction process replaces the mirrors and the beam splitters and, in comparison with optical interferometers, these diffraction processes can be separated either in space or in time. One can estimate the sensitivity of such an interferometer by calculating the area enclosed by the two atomic trajectories. The two atomic wavepackets are separated in velocity by the photon recoil $v_{\rm rec} = \hbar \keff/m$ of a few cm/s. The area is then simply the product $\hbar \keff L T$ where $T$ is the time left before the two atomic waves are redirected and $L$ the distance between the first and the last laser pulse.

Why is atom interferometry so powerful? In an optical interferometer, the electromagnetic waves travel very quickly at the speed of the light. In an atom interferometer, the atomic waves, traveling at much slower speeds, can spend a much longer time being interrogated. Thus, atom interferometers are more sensitive to their environment than their optical counterpart---up to $10^{11}$ times for the same interferometer area and signal-to-noise ratio in the case, for instance, of an Sagnac-effect based gyroscope. Thus, even if the interferometer area and SNR are favorable to photonics interferometers, typically a 1 cm$^2$/1 m$^2$ ratio for the area and a SNR ratio smaller than 1:1000, atom interferometers are still exhibiting extreme sensitivities. In addition, in addition to the precise control of the interferometer scale factor, which varies only with time and laser wavelength, many systematic effects such as, for example, laser wavefront distortion \cite{PhysRevA.92.013616} or magnetic fields inhomogeneities\cite{Peters2001,Gillot2014}, can be characterized with great accuracy, thus leading to minimal bias instabilities and drifts.

This accuracy and sensitivity can be used for very precise measurements such as, for instance, the effect of electric or magnetic fields on atoms, the mass of an atom (for tests of certain fundamental laws of physics), decoherence and collision effects (index of refraction for atomic waves) and inertial effects such as the acceleration of gravity (with possible applications in mineral prospecting) or the rotation/acceleration undergone by the interferometer (the atom interferometer then becomes an inertial sensor). Apart from potential applications in navigation, this sensitivity to inertial effects can be used for testing fundamental physics and is expected to reach its full potential only in Space.

\section{Inertial quantum sensors}

When the atoms in the interferometer are subject to acceleration or rotation, their velocity along their trajectory is modified. This results in a variation of the atomic de Broglie wavelength, which itself leads to a dephasing between the two interferometer arms that shifts the interference pattern in each output port \cite{Storey1994}. The phase shift can be strong, \eg $\sim 10^7$ rad for a rubidium interferometer with $T = 250$ ms. Consequently, these interferometers can be more sensitive to their environment than their optical counterpart.

In 1991, proof-of-principle atom interferometers measured rotations \cite{Riehle1991} and accelerations \cite{Kasevich1992}. In the following years, many theoretical and experimental work was carried out to investigate new types of inertial sensors based on atom interferometry. Recent work has demonstrated that rotations and accelerations can be monitored with extremely high accuracy and sensitivity \cite{Barrett2014b}.

\subsection{Sensitivity to rotation}

When the three light pulses constituting the Mach-Zehnder atom interferometer are spatially separated with common orientations $\bm{k}_{\rm eff}$ (usually the atomic velocity $\bm{v}$ is perpendicular to the direction of the laser beams $\bm{k}_{\rm eff}$), the interferometer is sensitive to rotations, as in the Sagnac geometry for light interferometers. It is straightforward to show that for a Sagnac loop enclosing area $\bm{A} = \bm{k}_{\rm eff} \times \bm{v} T^2$, a rotation $\bm{\Omega}$ produces a phase shift (to first order in $\Omega$):
\be
  \label{eq:sagnacphase}
  \Delta\phi_{\rm rot} = \frac{4\pi}{\lambda v}~\bm{A} \cdot \bm{\Omega}.
\ee
Gyroscopes built on this principle have achieved performance levels in the laboratory which compare favorably with state-of-the-art optical gyroscopes \cite{Lefevre2014} at the level of a few $10^{-9}$ rad/s after 1000 seconds of integration \cite{Barrett2014b}.

\begin{figure}[!tb]
  \centering
  \includegraphics[width=0.9\textwidth]{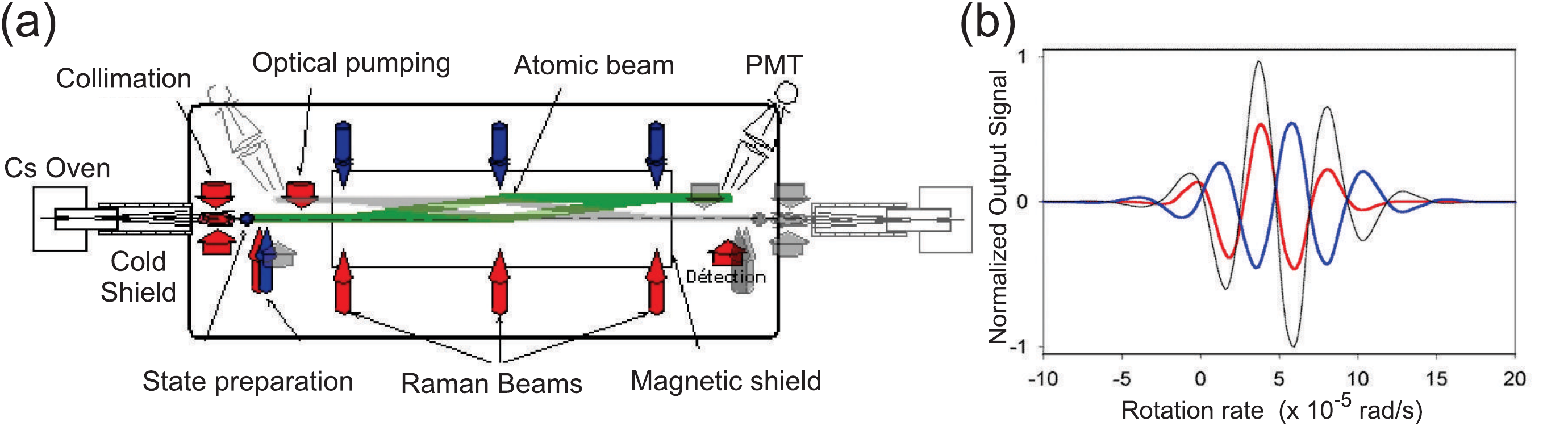}
  \caption{(a) General scheme of a matter-wave Sagnac interferometer (taken from \Ref \cite{Gustavson1997}). Atoms are launched from the two sides of the interferometer with velocities $\pm\bm{v}$. They interact with the same three beam splitters. (b) Individual signals from each interferometers (gray lines), and difference corresponding to a pure rotation signal (black line) versus rotation rate.}
  \label{fig:DoubleInt}
\end{figure}

\subsection{Sensitivity to acceleration}

If the platform containing the laser beams accelerates, or if the atoms are subject to a gravitational acceleration, the phase shift then contains the acceleration term
\be
  \label{eq:accphase}
  \Delta\phi_{\rm acc} = \bm{k}_{\rm eff} \cdot \bm{g} T^2.
\ee
For a stationary interferometer, with the laser beams vertically directed, this phase shift measures the acceleration due to gravity $\bm{g}$. Remarkably, ppb-level agreement has been achieved between the output of an atom-interferometric gravimeter and a conventional, ``falling-corner-cube'' gravimeter \cite{Merlet2010}. Such devices are used to measure gravity with high accuracy in a Watt balance that serves as the redefinition of the kilogram \cite{Merlet2008}.

Due to the stability of their acceleration outputs, a pair of such light pulse accelerometers is well-suited to gravity gradient instrumentation \cite{Snadden1998}. The basic idea is to simultaneously create two spatially separated interferometers using a common laser beam. In this way, technical acceleration noise of the measurement platform is a common-mode noise source which leads to near identical phase shifts in each accelerometer. On the other hand, a gravity gradient along the measurement axis results in a residual differential phase shift. This configuration has been used to measure the gravity gradient of the Earth, as well as the gravity gradient associated with nearby mass distributions. Laboratory gravity gradiometers have achieved resolutions below 1 E (where 1 E = $10^{-9}$ s$^{-2}$) and were used to measure the gravitational constant $G$ \cite{Bertoldi2006, Fixler2007, Rosi2014}. In addition, configurations similar to those for measuring gravity gradients can be used for gravitational wave detection \cite{Canuel2014, Geiger2015, Chaibi2016}.

\subsection{Quantum navigation system}

\begin{figure}[!tb]
  \centering
  \includegraphics[width=0.9\textwidth]{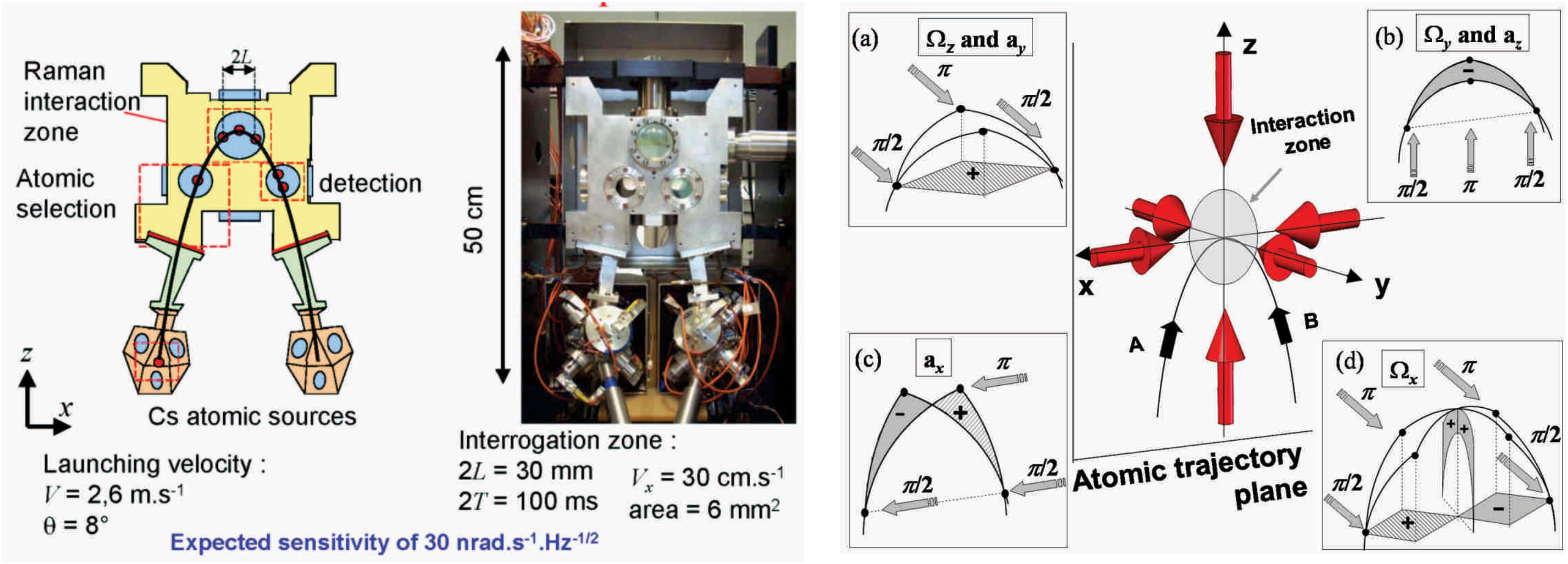}
  \caption{Six-axis inertial sensor: cold atomic clouds are launched on a parabolic trajectory, and interact with the lasers at the top of the parabola. Four interferometer configurations (a)-(d) give access to the three components of rotation and the three components of acceleration.}
  \label{fig:Navigation}
\end{figure}

The navigation problem can be simply stated by the following question: How do we determine an object's position and orientation as a function of time? In the 20$^{\rm th}$ century, investigations of this problem have lead to the development of exquisitely refined hardware, positioning systems and navigation algorithms. Today we take for granted that a hand-held GPS receiver can be used to obtain meter-level position determination. When GPS is unavailable (for example, when satellites are not in view, or the GPS signal is too weak), position determination becomes much less accurate. In this case, stand-alone ``black-box'' inertial measurements units (IMUs), comprised of a three orthogonal gyroscopes and accelerometers, are used to infer position changes by integrating the outputs of these inertial force sensors. State-of-the-art navigation-grade IMUs have position drift errors of a few kilometers per hour of navigation time---significantly worse than the GPS solution.

Light-pulse atom sensors appear well suited to this challenge. In general, both rotation and acceleration terms are present in the sensor outputs. For navigation applications, the rotation response needs to be isolated from the acceleration response. In practice, this is accomplished by using multiple atom sources and laser beam propagation axes. It is interesting to note that the same apparatus is thus capable of providing both rotation and acceleration outputs (see \Fig \ref{fig:Navigation})---a significant benefit for navigation applications which require the rotation rate and acceleration for three mutually orthogonal axes.

\begin{figure}[!tb]
  \centering
  \includegraphics[width=0.6\textwidth]{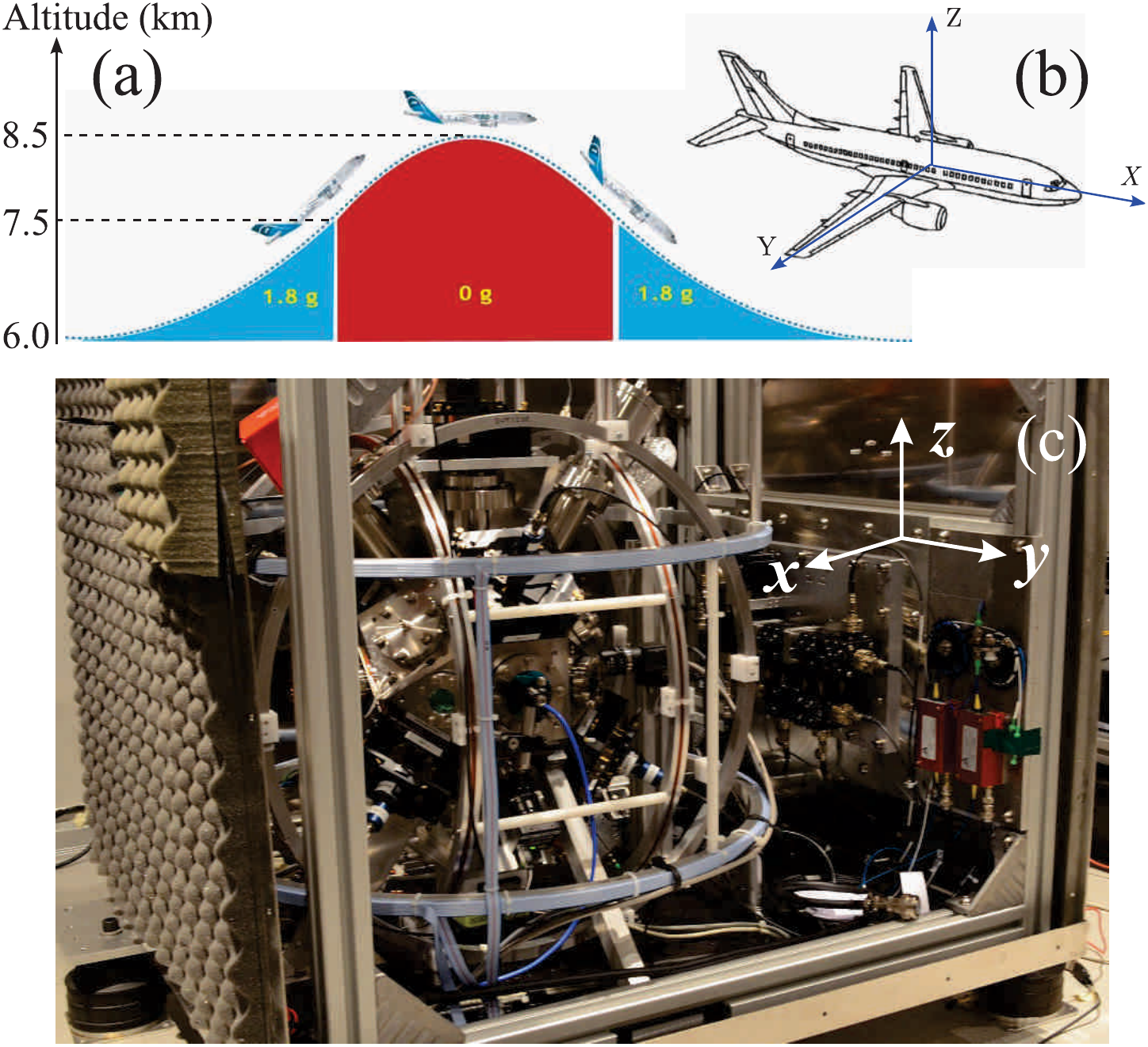}
  \caption{(a) Basic trajectory of the Novespace Zero-G aircraft during parabolic flight. (b) Coordinate system onboard the aircraft. (c) The science chamber mounted onboard the aircraft. Samples of $^{87}$Rb and $^{39}$K are laser-cooled in a vapor-loaded magneto-optical trap contained within a titanium vacuum system and enclosed by a $\mu$-metal magnetic shield. Raman beams are aligned either along the $y$- or the $z$-axis of the aircraft.}
  \label{fig:ICE-Setup}
\end{figure}

Recently, we have operated an inertial quantum sensor based on atom interferometry onboard an aircraft (the Novespace 0g airbus, see \Fig \ref{fig:ICE-Setup}) during both steady flight (standard gravity, $1g$) and parabolic flight (microgravity, $0g$). In steady flight, the inertial measurements performed by our instrument showed how a matter-wave sensor can achieve a resolution level 300 times below the aircraft's acceleration level \cite{Geiger2011}---a suitable noise base for inertial navigation.

An ideal feature of our experiment is the use of telecom-based laser sources that provide high stability in frequency and power in a compact and integrated setup. We use a cloud of about $3 \times 10^7$ $^{87}$Rb atoms laser-cooled to $\sim 5$ $\mu$K. After selecting a magnetic-field-insensitive ($m_F = 0$) Zeeman sub-level, we apply a velocity-selective Raman light pulse \cite{Kasevich1991a, Moler1992} that keeps about $10^6$ atoms at a temperature of 300 nK along the Raman beams. In a first experiment, the Raman laser beams were aligned along the direction of the plane's wings ($y$-axis in \Fig \ref{fig:ICE-Setup}) and retro-reflected by a mirror attached to the aircraft's structure, thus following the motion of the aircraft. The acceleration measurement process can be pictured as tracking successive positions of the free-falling atoms with the pair of Raman lasers, and the resulting atomic phase shift $\Phi$ is the difference between the phase of the two Raman lasers at the atom's successive classical positions, with respect to the retro-reflecting mirror \cite{Storey1994}. This phase simply relates to the distance between the atoms and the reference mirror. Thus, our measurement is equivalent to recording the relative acceleration $a$ of the mirror (along the Raman beam axis) during the interferometer duration. The sinusoidal signal at the output of the interferometer is modulated by the acceleration-induced atomic phase shift $\Phi = \keff a T^2$, where $\keff \simeq 4\pi/780$ nm is the effective wavevector of the Raman light, and $T$ is the time between the light pulses, with $a$ the time-dependent acceleration of the aircraft.

\begin{figure}[!tb]
  \centering
  \includegraphics[width=0.8\textwidth]{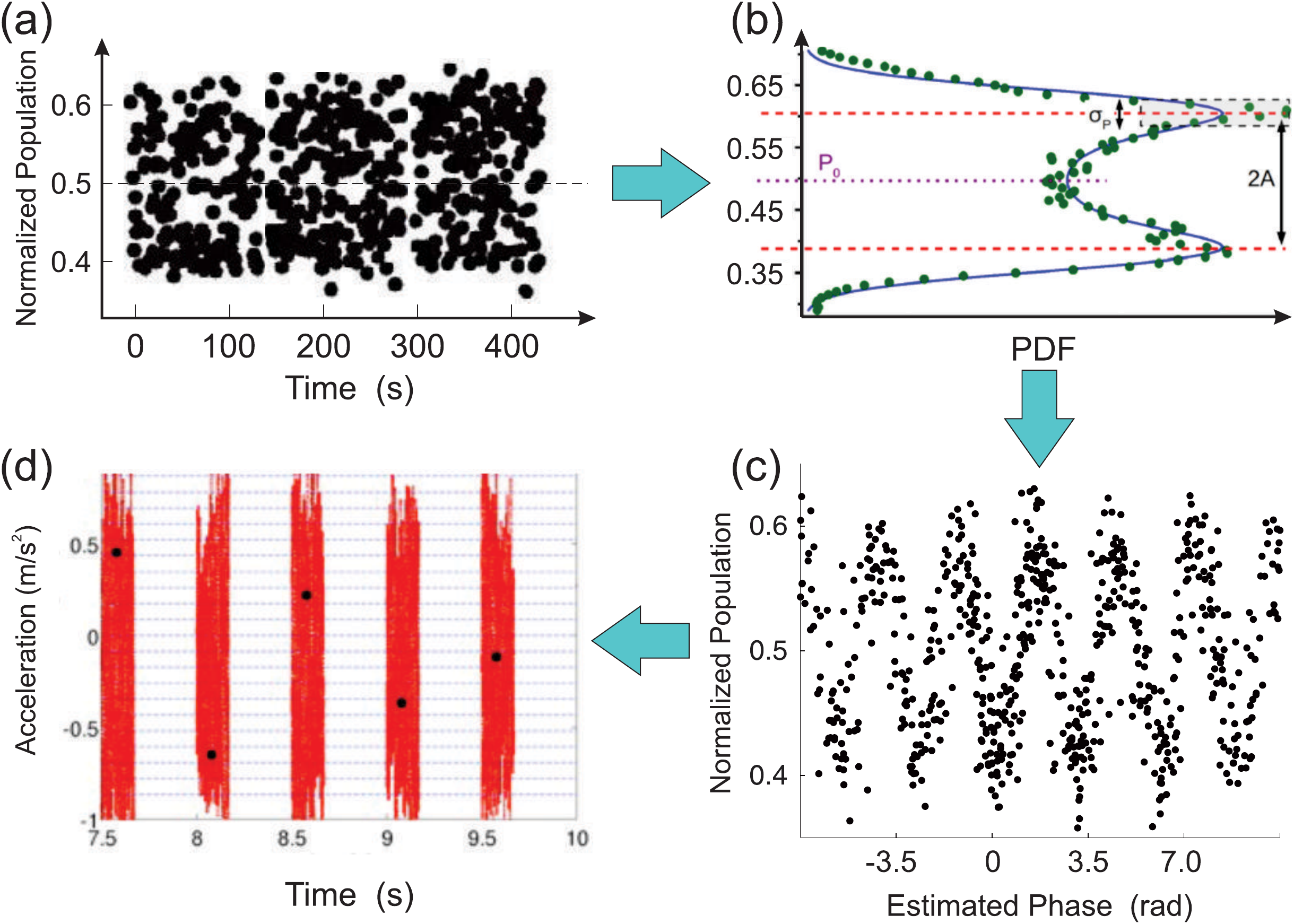}
  \caption{(a) The signal of the atom interferometer recorded during the flight cannot by itself be used because the interference fringes are smeared out by the acceleration noise. The statistics of the AI output (b) still shows a characteristic probability density function (PDF)---confirming that interference fringes can be reconstructed by correlating the low sensitivity output of a mechanical accelerometer with the atom interferometer (c). The value provided by the atom interferometer can then be used to refine the acceleration measurement and increase the sensitivity of the navigation system (d).}
  \label{fig:Correlation}
\end{figure}

In the aircraft, the acceleration fluctuates over time by $\sigma_a \simeq 0.5$ m/s$^2$ (1 standard deviation), and is at least four orders of magnitude greater than the typical signal variations recorded by laboratory-based matter-wave inertial sensors. For this reason, the signal recorded by the atom interferometer first appears as random. To retrieve the information of our interferometer, we use a mechanical accelerometer (MA) fixed on the retro-reflecting mirror and correlate the measurements between the MA and the atom interferometer \cite{Merlet2009, Geiger2011, Barrett2014a}. We use the signal continuously recorded by the MA to estimate the mean acceleration $a(t)$ which is expected to be measured by the atom interferometer at time $t$. Plotting the atomic measurements $P(t)$ versus $a(t)$ reveals clear sinusoidal correlations between the mechanical sensors and the atom interferometer (see \Fig \ref{fig:Correlation}). This demonstrates that the atom interferometer truly holds information about the mirror acceleration. Using the information of the MA to locate the interferometer fringe, and the sensitivity of the atomic measurement, our hybrid sensor was able to surpass the dynamic range of stand-alone atom-based accelerometers ($\pm \pi/k_{\rm eff}T^2$), and measure large accelerations ($\pm 0.5$ m/s$^2$) and at high resolution ($2 \times 10^{-3}$ m/s$^2$ per shot) \cite{Geiger2011}.

\section{Fundamental physics and quantum tests of weak equivalence principle}

The intrinsically high sensitivity and stability of atom interferometers enables access to many types of precision tests of fundamental physics. They allow, for instance, to test atomic and molecular physics properties at a level yet unachieved. For example, by applying a perturbation on one of the two atomic paths inside the interferometer, the electric polarizability of atoms or molecules \cite{Cronin2015b}, as well as the index of refraction of gases for atomic waves or topological phases can be measured \cite{Lepoutre2013}. As for atomic clocks, atom interferometers can also be used to measure fundamental constants at a very high level of accuracy. Present efforts include the measurement of the fine constant structure $\alpha$ \cite{Bouchendira2011}, the gravitational constant $G$ \cite{Bertoldi2006, Fixler2007, Rosi2014} and the definition of the kilogram.

There are also several interesting experimental tests of General Relativity (GR), motivated by alternatives to Einstein's theory, that could be within reach with atom interferometers \cite{Dimopoulos2007}. For example, the cosmological constant problem suggests that our understanding of GR is incomplete, motivating a modification of gravity at large distances. Atom interferometry can lead to high precision laboratory tests of GR, such as the test of the Einstein's equivalence principle (in the form of the UFF, as discussed below) or the detection of gravitational waves at low frequency. There are two main reasons for this. Atomic physics experiments can reach incredibly high accuracy (for example, clock synchronization at levels of a few $10^{-17}$ \cite{Lisdat2015} and $10^{-18}$ \cite{Ushijima2015} of relative accuracy), and have several control parameters which, for instance, allow one to isolate and measure individual relativistic terms by using their scalings with these parameters.

Precise tests of the UFF with matter waves are of key importance to understanding gravity at the quantum scale. Such tests use two atom interferometers that measure the relative acceleration between two atomic species in free fall with the Earth's gravitational potential. The ICE experiment \cite{Barrett2015} is designed to generate interferometer signals from laser-cooled samples of $^{39}$K and $^{87}$Rb onboard the Novespace Zero-g aircraft. While in parabolic flight, the experiment is in free fall and this microgravity environment should enable interrogation times on the order of 10 s. Since the sensitivity of atom interferometers to acceleration scales as the square of the interrogation interval $T$, measurements on this timescale could theoretically detect changes in acceleration at the level of $10^{-11}\,g$. In order to laser-cool $^{39}$K and $^{87}$Rb, two independent laser sources are required at 767 and 780 nm. These wavelengths are generated by frequency-doubling the output of two C-band telecom lasers operating at 1560 and 1534 nm \cite{Menoret2011}. By using these fiber-based lasers the system is largely insensitive to optical misalignment due to vibrations or structural deformation. The frequency of both lasers is stabilized by locking them to an optical frequency comb that operates over both telecom wavelengths. After amplification and frequency-doubling, the light at 767 and 780 nm passes through a series of acousto-optic modulators in free space (which act as frequency shifters and high-speed optical switches) before being sent through optical fibers toward the vacuum chamber.

We recently made simultaneous acceleration measurements with the two atomic species in microgravity. Since the two interferometer signals originate from atomic sources that occupy the same space, many systematic effects related to a precise test of the UFF can be eliminated. Our experiment can generate sources of cold $^{87}$Rb and $^{39}$K samples at temperatures of $\sim 2$ $\mu$K and $\sim 18$ $\mu$K, respectively, each with approximately $10^8$ atoms. The Raman beams at 780 nm and 767 nm are combined on the same optics before being aligned through the atomic cloud and retro-reflected off of a reference mirror. In this way, mirror vibrations are common to both interferometer signals \cite{Varoquaux2009, Bonnin2015}, and many sources of measurement noise can be rejected to a high degree. In addition, a high-sensitivity mechanical accelerometer (Colibrys SF3600) is attached to this mirror and its signal is combined with the output of the two interferometers to further reduce noise due to low-frequency vibrations and mirror drift. This technique is effective at removing phase noise even if no vibration isolation system is used.

\section{Atom lasers, quantum phase locks and sub-shot-noise interferometry}

Laser cooling, recognized by the 1997 Nobel prize in physics \cite{Chu1997}, made it possible to reach high atomic densities (\ie to observe a strong signal), while combining temperature of atom clouds very close to the absolute zero (1 microKelvin) and thus to reach regimes where the wave behavior of the atoms becomes significant. Nevertheless, the quest to reach higher sensitivity requires (i) reducing the velocity dispersion of the atomic sample \cite{Kovachy2015b} in order to allow for enhanced enclosed area via larger momentum splitting \cite{Kovachy2015}, (ii) increasing the interrogation time, or (iii) increasing the knowledge of the scaling factor, which depends directly on the initial velocity of the atoms and can be better controlled with more confined atomic sources.

After the first observation of Bose-Einstein condensation (BEC) in dilute atomic gases (which was awarded the Nobel prize in physics in 2001 \cite{Cornell2001}), the possibility of producing dense, \emph{ultra-cold} samples of atoms opened new prospects in our ability to engineer quantum sensors. Laboratories all around the world are now routinely producing evaporatively-cooled (below 100 nK) samples of atoms which, at a sufficiently high density, undergo a phase transition to quantum degeneracy. For a cloud of bosonic (integer spin) atoms, all the atoms accumulate in the same quantum state (the atom-optical analog of the laser effect in optics). Access to BECs and atom laser sources have brought major conceptual advances in atom interferometry \cite{Gross2010, VanZoest2010, Sugarbaker2013, Dickerson2013, Muntinga2013}, in a similar fashion to what lasers did for the field of optical interferometry. Nevertheless, the relative complexity of BEC production has pushed scientists to explore new techniques by using, for instance, optical traps \cite{Grimm2000}, atom chips \cite{Reichel2011}, or avoiding evaporation and relying solely on laser cooling \cite{Stellmer2013}.

\begin{figure}[!tb]
  \centering
  \subfigure{\includegraphics[width=0.4\textwidth]{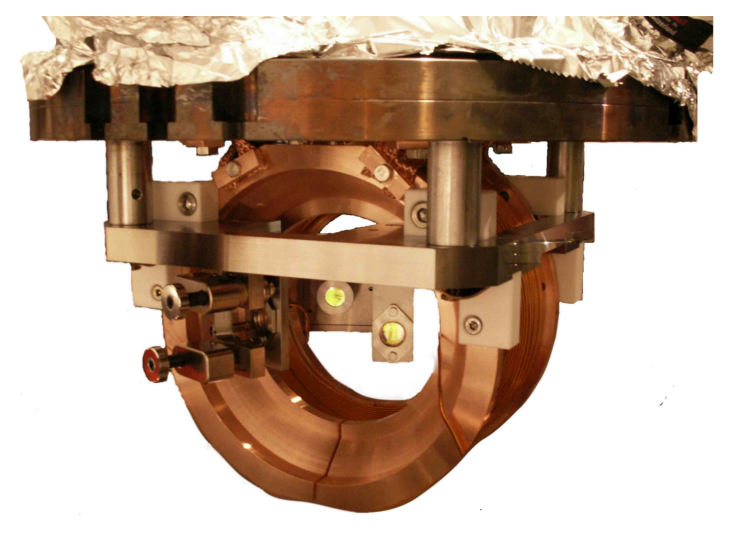}}
  \hspace{0.1cm}
  \subfigure{\includegraphics[width=0.4\textwidth]{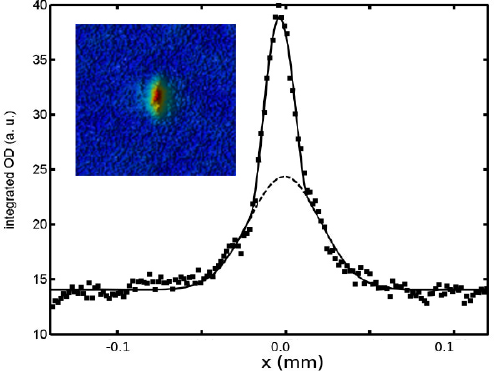}}
  \caption{(left) The bow-tie cavity mounted with the MOT coils on a metal flange; the assembly is held in vacuum and used to trap and cool Rb atoms until condensation. (right) A Bose-condensed sample of $5 \times 10^5$ Rb atoms obtained by evaporatively cooling atoms in the cavity-enhanced optical potential: in the inset, the optical density of the atomic ensemble obtained from an absorption image after 8 ms of time of flight; in the graph, the optical density integrated along the vertical direction shows a bimodal structure, with the inverted parabola accounting for the condensed atomic fraction.}
  \label{fig:becBiaro}
\end{figure}

In this context, we recently managed to produce a rubidium BEC in an optical cavity held in vacuum (\Fig \ref{fig:becBiaro}): the atoms from a magneto-optical trap are transferred to the dipole optical trap created by a telecom laser at 1560 nm coupled into the cavity. The system consists of a four-mirror cavity in a bow-tie configuration, with two arms crossing orthogonally at the center. As a result of cavity build-up, the intensity is much higher inside the cavity---enabling the design of a compact, low power, BEC production system. Since the transverse modes of the optical cavity are non-degenerate, it is also possible to selectively lock the trapping laser to such modes. By tuning the laser frequency and adopting suitable phase masks to optimize the coupling efficiency \cite{Bertoldi2010}, we are able to confine thermal atoms in the trap arrays formed by the high-order modes, and also dynamically switch the trap geometry by injecting two cavity modes with a variable intensity. This scheme could ultimately be used to realize compact arrays of simultaneous interferometers.

The resonator-based generation of BEC enables a drastic reduction of the power required for the laser source, and could constitute a key approach for transportable quantum sensors using condensed sources. The main drawback of this configuration is that the high sensitivity of the intra-cavity laser intensity to vibrations results in a significant heating for the trapped atoms. However, this mechanism could be strongly suppressed using a novel scheme that we developed to lock a laser to a cavity using a serrodyne modulation technique \cite{Kohlhaas2012}.

\begin{figure}[!tb]
  \centering
  \includegraphics[width=0.7\textwidth]{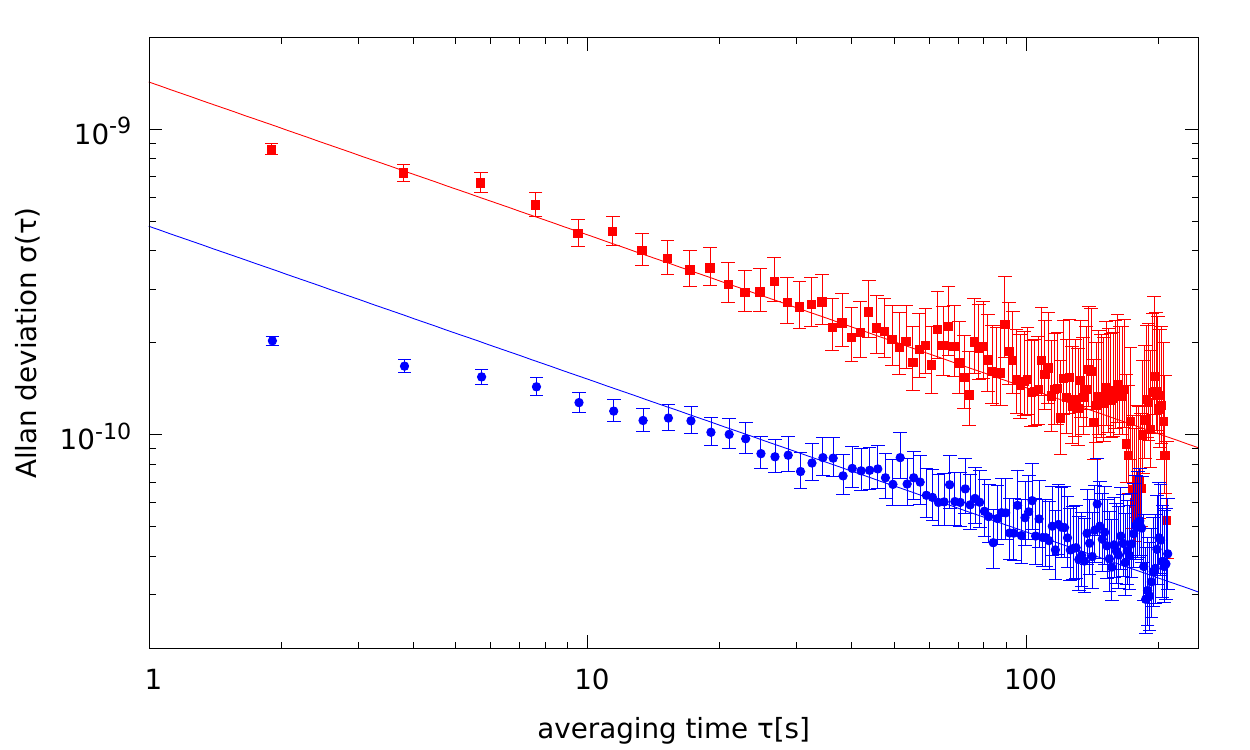}
  \caption{Comparison of the Allan deviation of a reference clock with $T = 1$ ms (red points) and that of a model clock implementing 9 successive interrogations with intermediate measurements of the relative phase and corrections, for a total interrogation time of 9 ms (blue points). Image taken from \Ref \cite{Kohlhaas2015}.}
  \label{fig:allanDev}
\end{figure}

Such a system, combined with improved atom interferometer detection, could be used to improve quantum sensors in many respects. For instance, a common issue in atom interferometry is represented by the coherence loss determined by the limited interval over which the interferometric phase can be unambiguously determined from the measurement of a population unbalance over the two output ports. This is the case with both atomic clocks, where the quality of the local oscillator used to interrogate the atomic coherence ultimately limits the measurement, and with accelerometers and gyroscopes based on matter-wave interference, where the fringe visibility is degraded by acceleration and rotation noise, respectively. Several solutions to extend the interrogation interval have been proposed for atomic clocks \cite{Rosenband2013, Borregaard2013} and demonstrated in atom interferometry based inertial sensing \cite{Sorrentino2012}; they use two or more ensembles interrogated simultaneously at different time scales or locations to increase the phase-wrap-free interval. With our system, we can measure the population imbalance in an atom interferometer with a minimal destructivity of the ensemble coherence. We have used this system to operate a feedback correction on the atomic state to protect its coherence against random noise \cite{Vanderbruggen2013}, and also to phase lock a microwave frequency chain to a coherent atomic state \cite{Shiga2012, Kohlhaas2015}. This scheme allowed us to increase the effective interrogation time in an atomic clock beyond the limit set by the quality of the local oscillator. In \Fig \ref{fig:allanDev}, an atomic clock implementing the phase lock method is compared to a clock using a standard Ramsey interrogation, and the performance obtained indicates a certain degree of correlation between the successive interrogations in the same clock cycle. The coherence preserving measurements and feedback scheme, which allowed us to demonstrate an effectively longer interrogation interval, could improve the sensitivity of optical atomic clocks bypassing the limitations set by the local oscillator adopted for the coherent manipulation. We plan to apply it to matter-wave-based inertial sensors to enhance fringe visibility at long interrogation times or in harsh environments.

Aside from technical difficulties, the sensitivity of an atom interferometer will be eventually limited by the quantum projection noise limit $\Delta \phi_{\rm min} = 1/\sqrt{N}$ for a finite number of uncorrelated detected particles $N$ \cite{Itano1993}. With the relative phases accumulated in the interferometer being a function of the interrogation time $T$, the ideal sensitivity will scale as $\sqrt{N}\,T^{\alpha}$, where $\alpha > 0$ and its value depends on the interferometer configuration. Using cold atomic sources can enable the enhancement of the enclosed area and the sensitivity by increasing $T$. On the other hand, the effect of quantum projection noise can be reduced by engineering the quantum state of the atomic ensemble entering the interferometer: entanglement can determine strong correlations in the projection process during the measurement, which can be exploited to increase phase resolution with the Heisenberg limit $\Delta \phi_{\rm min,H} = 1/N$ as fundamental lower bound. After the first experimental realizations of atomic states with uncertainties below the quantum projection limit \cite{Appel2009,Schleier-Smith2010,Riedel2010,Gross2010,Bucker2011,Lucke2011}, a noise reduction approaching -20 dB below the QPN has been recently reported \cite{Hosten2016, Cox2016}. Our heterodyne technique, when used for quantum non-demolition measurements on the atomic sample, could combine the enhancement given by correlating successive measurements on the atomic system with that related to implementing entanglement between the particles so as to reach Heisenberg-limited sensitivities at the output of the interferometer. The ultimate bound of this combined approach has been recently studied for atomic clocks \cite{Chabuda2016}.

\section{Conclusion}

The last few decades have advanced our ability to produce very reliably ultra-cold sample of atoms, and has open the possibility of new technologies based on the quantum manipulation of matter waves. While ``quantum technologies'' represent tomorrow's challenge in computing, communication or sensing, the manipulation of quantum states of matter with light has lead to a promising class of inertial sensors that are already nowadays exhibiting exquisite sensitivity to rotation and gravity. These quantum sensors are showing great promise for improving inertial navigation, or for passive sub-surface monitoring and prospecting. Furthermore, matter-wave interference and entanglement of quantum states also enable a new generation of fundamental physics tests, where laboratory experiments might reach the precision and accuracy of large astrophysical experiments, and where the foundation of our understanding of physics could be challenged.

What lies in the future? Developments in quantum sensors involve combinations of advanced atomic sources and clever beam-splitting elements such as high-flux, single-mode atomic sources and large momentum transfer atom optics. There are several promising avenues to realizing this goal. The pay-off: extremely high performing space-time sensors with applications ranging from tests of gravitation to advanced inertial sensing. Nevertheless, our quest for improving the sensitivity of ground-based atom interferometers will soon reach a limit imposed by gravity and by the requirements of ultra-high vacuum and a very well controlled environment. Current state-of-the-art experimental apparatuses allow for seconds of interrogation with 10 to 120 meters of free-fall \cite{VanZoest2010, Sugarbaker2013, Dickerson2013, Muntinga2013, Kovachy2015}. Space-based applications \footnote{See for example the Special Issue: Quantum Mechanics for Space Application: From Quantum Optics to Atom Optics and General Relativity, Appl. Phys. B, 84 (2006);}, currently under study, will enable physicists to increase even further the interrogation time, thereby increasing dramatically the sensitivity and accuracy of atom interferometers.

\section{Acknowledgments}

The work presented here is supported by the French national agencies CNES (Centre National d'Etudes Spatiales), l'Agence Nationale pour la Recherche, the D\'{e}l\'egation G\'{e}n\'{e}rale de l'Armement, the European Space Agency, IFRAF (Institut Francilien de Recherche sur les Atomes Froids), the European Metrology Research Programme (EMRP) (JRP-EXL01 QESOCAS), Laser and Photonics in Aquitaine (APLL-CLOCK, within ANR-10-IDEX-03-02). We would like to thank our collaborators B. Canuel, B. Battelier and A. Landragin. P. Bouyer thanks Conseil R\'{e}gional d'Aquitaine for the Excellence Chair and M. Kasevich who is involved in the BSAIL international laboratory.

\section*{References}

\bibliographystyle{iopart-num}
\bibliography{References}
\end{document}